\documentclass[12pt]{article}

\title{}
\author{}

\usepackage{url}

\usepackage[table,xcdraw]{xcolor}

\usepackage{graphicx}

\usepackage{caption}
\usepackage{subcaption}

\usepackage{amsmath} 
\usepackage{amsfonts}
\usepackage{nicefrac}

\usepackage{hyperref}
\hypersetup{
	colorlinks=true,
	linkcolor=blue,
	filecolor=blue,      
	urlcolor=blue,
	citecolor=blue,
}

\usepackage[a4paper, total={6in, 8in}]{geometry}

\usepackage{mhchem}
\usepackage{algorithm}
\usepackage{algpseudocode}

\usepackage[symbol]{footmisc}

\usepackage{multirow}

\begin{document}
	\begin{center}	
\sf {\Large {\bfseries Self-Supervised Adversarial Diffusion Models for Fast MRI Reconstruction}} \\  
		\vspace*{10mm}
		Mojtaba Safari$ ^{1} $, Zach Eidex$ ^{1} $, Shaoyan Pan$ ^{1} $, Richard L.J. Qiu$ ^{1} $, Xiaofeng Yang, PhD$ ^{1,\ddagger} $ \\
	{$ ^1 $Department of Radiation Oncology and Winship Cancer Institute, Emory University, Atlanta, GA 30322, United States of America\\$ ^\ddagger  $Corresponding Author}
	\vspace{5mm}\\
\end{center}

Author to whom correspondence should be addressed. email: \url{xiaofeng.yang@emory.edu} \\


\begin{abstract}
	\noindent \textbf{Background:} Magnetic resonance imaging (MRI) offers excellent soft tissue contrast essential for diagnosis and treatment, but its long acquisition times can cause patient discomfort and motion artifacts. \\ 
	
	\noindent\textbf{Purpose:} To propose a self-supervised deep learning-based compressed sensing MRI (DL-based CS-MRI) method named ``Self-Supervised Adversarial Diffusion for MRI Accelerated Reconstruction (SSAD-MRI)'' to accelerate data acquisition without requiring fully sampled datasets.   \\	
	
	\noindent\textbf{Materials and Methods:} We used the fastMRI multi-coil brain axial \ce{T2}-weighted (\ce{T2}-w) dataset from 1,376 cases and single-coil brain quantitative magnetization prepared 2 rapid acquisition gradient echoes (MP2RAGE) \ce{T1} maps from 318 cases to train and test our model. Robustness against domain shift was evaluated using two out-of-distribution (OOD) datasets: multi-coil brain axial postcontrast \ce{T1}-weighted (\ce{T1}c) dataset from 50 cases and axial T1-weighted (T1-w) dataset from 50 patients. Data were retrospectively subsampled at acceleration rates $ R \in \{2\times, 4\times, 8\times\} $. SSAD-MRI partitions a random sampling pattern into two disjoint sets, ensuring data consistency during training. We compared our method with ReconFormer Transformer and SS-MRI, assessing performance using normalized mean squared error (NMSE), peak signal-to-noise ratio (PSNR), and structural similarity index (SSIM). Statistical tests included one-way analysis of variance (ANOVA) and multi-comparison Tukey’s Honesty Significant Difference (HSD) tests.\\

	\noindent\textbf{Results:} SSAD-MRI preserved fine structures and brain abnormalities visually better than comparative methods at $ R=8\times $ for both multi-coil and single-coil datasets. It achieved the lowest NMSE at $ R \in \{4\times, 8\times\} $, and the highest PSNR and SSIM values at all acceleration rates for the multi-coil dataset. Similar trends were observed for the single-coil dataset, though SSIM values were comparable to ReconFormer at $ R \in \{2\times, 8\times\} $. These results were further confirmed by the voxel-wise correlation scatter plots. OOD results showed significant (\textit{p} $ \ll 10^{-5}$) improvements in undersampled image quality after reconstruction.\\

	\noindent\textbf{Conclusions:} SSAD-MRI successfully reconstructs fully sampled images without utilizing them in the training step, potentially reducing imaging costs and enhancing image quality crucial for diagnosis and treatment.
	
\end{abstract}

\textbf{\textit{keywords:}} \textit{k}-space sampling, fastMRI, accelerated MRI, reconstruction, adaptive partitioning

\section{Introduction}

Magnetic Resonance Imaging (MRI) provides excellent soft tissue contrast, playing a vital role in diagnosis, treatment, and follow-up. However, prolonged acquisition times can lead to patient discomfort and increase the likelihood of motion artifacts, which compromise image quality. Additionally, the requirement for highly sampled k-space data to achieve high-resolution MRI images inherently extends acquisition times, reducing imaging throughput and limiting patient access to MRI services. This issue is particularly significant in low- and middle-income countries, where the recent Lancet Oncology Commission highlighted severe shortages of MRI and other medical imaging devices, potentially resulting in 2.5 million preventable deaths worldwide~\cite{hricak2021medical, liu2021low}. Globally, only about seven MRI scanners are installed per million people, largely due to the high costs of installation, operation, and maintenance. Therefore, developing techniques that can accelerate MRI acquisition without sacrificing image quality is crucial for improving accessibility, reducing operational costs, and enhancing patient care worldwide~\cite{safari2024unsupervised}.

The MRI acquisition can be accelerated by reducing the sampled \textit{k}-space data, but this is limited by the Nyquist criteria. Compressed sensing (CS) and parallel imaging (PI) techniques aim to recover fully sampled images from under-sampled images by exploiting data in a sparse transformed space and redundant data acquired using uncorrelated radiofrequency coils, respectively~\cite{haldar2010compressed,deshmane2012parallel}. However, at high acceleration rates, PI and CS methods suffer from noise amplification~\cite{otazo2010combination} and residual artifacts~\cite{chang2012nonlinear}, respectively.

Deep learning (DL) algorithms have been extensively used to reconstruct accelerated high-resolution MRI images. DL-based compressed sensing MRI (CS-MRI) approaches often blend data-driven learning with physics-guided modeling~\cite{safari2024fast}. Data-driven methods focus on learning mappings from under-sampled k-space data to fully-sampled images or \textit{k}-space representations~\cite{jin2023improving,cao2024high,narnhofer2019inverse, eidex2024temporal}, while physics-guided methods incorporate knowledge of the MRI acquisition process, such as coil sensitivity profiles and under-sampling patterns, to enforce data consistency and address the ill-posed nature of the inverse problem.

Many state-of-the-art methods integrate both approaches, leveraging the strengths of data-driven learning and physics-based constraints. For instance, unrolled optimization methods~\cite{kofler2023deep, qu2024radial,safari2024fastmri} mimic iterative reconstruction algorithms by alternating between data consistency enforcement and learned regularization in a fixed number of iterations. These methods often incorporate data consistency layers (DC layers) within deep neural networks to ensure that the reconstructed images adhere to the acquired k-space measurements~\cite{gungor2023adaptive,terpstra2023accelerated,schlemper2017deep}.

Classic works such as Deep ADMM-Net~\cite{sun2016deep} and the variational network~\cite{hammernik2018learning} exemplify this integration by unrolling iterative algorithms and embedding physics-based constraints within DL frameworks. Similarly, approaches like deep density priors~\cite{tezcan2018mr} and deep Bayesian estimation~\cite{luo2020mri,luo2023bayesian} combine data-driven learning with probabilistic modeling to capture complex image priors while respecting the underlying physics of MRI acquisition.

These methods are typically trained under supervised frameworks where the reference fully sampled images were utilized to train a model. However, obtaining the fully sampled images might be impractical in imaging scenarios such as cardiovascular MRI due to excessive involuntary movements, or diffusion MRI with echo planar imaging due to quick \ce{T2^*} signal decay~\cite{setsompop2013pushing}. Additionally, acquiring high-resolution anatomical brain MRI images can be prohibitively long.

In this study, we propose an self-supervised adversarial diffusion model to reconstruct fully sampled images without requiring them in the training step. Our proposed SSAD-MRI model is based on an adversarial mapper to reconstruct fully sampled MRI images. Our proposed method was evaluated using both single-coil and multi-coil MRI data, as well as two out-of-distribution (OOD) datasets. Our method leverages a recently proposed ReconFormer transformer~\cite{guo2023reconformer} as a generator and is compared with two state-of-the-art models. Our contributions are as follows:

\begin{itemize}
	\item To our knowledge, SSAD-MRI is the first study proposing a self-supervised method using an adversarial mapper.
	\item The proposed method performs the backward diffusion process in smaller steps that improve sampling efficiency.
	\item The proposed method's robustness against domain shift was evaluated at the test time,
	\item To our knowledge, It is the first self-supervised method aimed at reconstructing fully sampled quantitative magnetization prepared 2 rapid acquisition gradient echoes (MP2RAGE) \ce{T1} map,
\end{itemize}

\section{Materials and Methods}\label{sec:matrial}
\subsection{Compressed sensing MRI}\label{subsec:csmri}
Let $ Y \in \mathbb{C}^N $ denote the observed subsampled \textit{k}-space measurement corresponding to the subsampled image $ y \in \mathbb{C}^N $, while $ x\in \mathbb{C}^M $ represents the unobserved fully sampled data. Hereafter, we adopt the convention that lowercase letters denote data in the image domain, whereas uppercase letters represent data in the Fourier (\textit{k}-space) domain. The compressed sensing formulation is expressed as follows:

\begin{equation}\label{eq:csmri_01}
	\centering
	Y = \mathcal{A}_\Omega x + \delta
\end{equation}
where $ \delta \in \mathbb{C}^N $ is the additive acquisition noise and $ \mathcal{A}_\Omega \in \mathbb{C}^{N \times M} \to \mathbb{C}^N  $ give $ N \ll M $ represents the encoding operator. $ \mathcal{A}_\Omega $ is composed of a coil sensitivity map $ \mathcal{S} $, a Fourier transform $ \mathcal{F} $, and a sampling map with the specified pattern $ \Omega $ controlling the acceleration rate (R). The mathematical expression for the encoding operator is $ \mathcal{A}_\Omega = \Omega \odot \mathcal{F} \odot \mathcal{S}$. The MRI reconstruction is formulated as an unconstrained optimization problem as follows~\cite{safari2024fastmri, schlemper2017deep, yaman2020self},

\begin{equation}\label{eq:csmri_02}
	\centering
	\underset{x, \varphi}{\arg\min} \parallel x - f_{\text{SSAD-MRI}} (y \mid \varphi)\parallel_2^2 + \lambda \parallel Y - \mathcal{A}_\Omega x \parallel_2^2
\end{equation}
where $ f_{\text{SSAD-MRI}} $ represents our proposed model parametrized by $ \varphi $. The first and second terms represent the regularization and data consistency, and $ \lambda > 0 $ is a scalar regularization weight that balances between the data consistency and regularization terms.

\subsection{Diffusion model}
The diffusion model inspired by nonequilibrium thermodynamics aims to approximate complex and intractable distributions with a tractable one like normal Gaussian~\cite{sohl2015deep}. It consists of two process: the forward process and the reverse process.

\subsubsection*{Forward process:} 
The forward process utilizes a noise scheduler to add Gaussian noise to the noise-free $ y_0 $ using a first-order Markov process $ q(y_t|y_{t-1}) $ in a large number of steps $ T $, eventually converting $ y_T $ to a normal multivariate Gaussian $y_T \sim \mathcal{N}(y_T ; \pmb{0}, \mathbf{I}) $. This step employs the first-order Markov process to calculate $ q(y_t | y_{t-1}) $ as follows~\cite{ho2020denoising}: 

\begin{equation}\label{eq:forward_encoder_01}
	\centering
	\begin{aligned}
		&y_t = \sqrt{1 - \beta_t} y_{t-1} + \sqrt{\beta_t} \epsilon\\
		&q(y_t \mid y_{t-1}) = \mathcal{N}(y_t ; \sqrt{1 - \beta_t} y_{t-1}, \beta_t \mathbf{I} )
	\end{aligned}
\end{equation}
where $ \epsilon \sim \mathcal{N}(\epsilon ; \pmb{0}, \mathbf{I}) $ and $ \beta_t \in (0, 1)$ with $ \beta_1 = 10^{-4} $ is the noise variance. Assuming additive Gaussian noise, sampling $ y_t $ in an arbitrary step $ t $ can be calculated in closed form as follows:

\begin{equation}\label{eq:sampling_qt_from_q0}
	\centering
	q(y_t \mid y_0) = \mathcal{N}(y_t ; \sqrt{\bar{\alpha}_t} y_0, (1 - \bar{\alpha}_t) \mathbf{I})
\end{equation} 
where $ \alpha_t = 1 - \beta_t $ and $ \bar{\alpha}_t = \prod_{s=1}^{t} \alpha_s$.

\subsubsection*{Reverse process:} 
The reverse process gradually learns to remove the added noise in $ y_T $ to recover noise-free $ y_0 $. This process of training a network $ p_\varphi $ to generate $ y_0 $ from Gaussian noise $ y_T $. The reverse process will follow the forward process trajectories but in the reverse direction for small $ \beta $ values as follows~\cite{sohl2015deep, luo2022understanding, chan2024tutorial}:

\begin{equation}\label{eq:kl_loss_01}
	\centering
	\underset{\varphi}{\arg\min} \sum_{t = 2}^{T} \mathbb{E}_{q(y_t \mid y_0)}\left[ D_{\mathbb{KL}}\left( q(y_{t-1} \mid y_{t}, y_0)\parallel p_\varphi (y_{t - 1} \mid y_{t}) \right) \right]
\end{equation}
This is a denoising matching term where it learns the desired denoising transition step $ p_\varphi (y_{t - 1} \mid y_{t}) $ as an approximator to tractable, ground-truth denoising transition step $ q(y_{t-1} \mid y_{t}, y_0) $ given in (\ref{eq:qyt_1fromy_t_01}), where it is modeled as a Gaussian. 

\begin{equation}\label{eq:qyt_1fromy_t_01}
	\begin{aligned}
		&q(y_{t-1} \mid y_{t}, y_0)  = \mathcal{N}\left( y_{t-1}; \mu_q(y_t,y_0), \sigma_q(t) \mathbf{I}  \right)\\
		& \mu_q(y_t,y_0) = \dfrac{\sqrt{\alpha_t} (1 - \bar{\alpha}_{t-1}) y_t + \sqrt{\bar{\alpha}_{t-1}}y_0}{1-\bar{\alpha}_t}\\
		&\sigma_q(t) = \dfrac{(1 - {\alpha}_t) (1 - \bar{\alpha}_{t-1})}{1 - \bar{\alpha}_t}
	\end{aligned}
\end{equation}

Furthermore, all the $ \alpha $ terms are frozen at each timestep, it was shown that the loss function given in (\ref{eq:kl_loss_01}) becomes as follows:

\begin{equation}\label{eq:kl_loss_02}
	\centering
		\text{\L{}} =\underset{\varphi}{\arg\min} \dfrac{1}{2 \sigma_q^2(t)}\left[\parallel \mu_q(y_t, y_0) -  \mu_\varphi(y_t) \parallel_2^2\right]
\end{equation}
where $ \mu_\varphi(y_t) $ is the estimated average recovered image as follows:

\begin{equation}\label{eq:mu_phi_v02}
	\mu_\varphi(y_t) = \dfrac{\sqrt{\alpha_t} (1 - \bar{\alpha}_{t-1}) y_t + \sqrt{\bar{\alpha}_{t-1}}y_\varphi(y_t)}{1-\bar{\alpha}_t}
\end{equation}
where $ y_\varphi(y_t) $ is parameterized by our DL model to recover $ y_0 $ from noisy image $ y_t $ at a given step $ t $.

We employed an adversarial mapper to implicitly capture the conditional distribution for the reverse process steps. The generator $ f_{\text{SSAD-MRI}} $ is used to sample $ \hat{y}_t \sim p_\varphi (y_t \mid y_{t + k}) $. At the same time, a discriminator $ \mathcal{D}_\theta $ differentiates between samples $ \hat{y}_t $ and actual sample $ y_t $ sampled from the true denoising distribution $ q(y_t \mid  y_{t + k}) $. Our discriminator was coupled with a gradient penalty to improve learning~\cite{gungor2023adaptive,chan2024tutorial}:

To improve the efficiency of modeling the reverse diffusion process, we propose an adversarial mapper that implicitly learns the conditional distribution between time steps without the need for an explicit formulation. This mapper consists of a generator and a discriminator that train simultaneously. The generator, denoted as $f_{\text{SSAD-MRI}}$, is designed to estimate the conditional distribution $p_\varphi(y_t \mid y_{t+k})$ by generating samples $\hat{y}_t$ that closely approximate the true intermediate states $y_t$, given the future states $y_{t+k}$. This process effectively facilitates the mapping from $y_{t+k}$ back to $y_t$.

The discriminator, represented as $D_\theta$, is trained to differentiate between the outputs produced by the generator, denoted as $\hat{y}_t$, and the true samples $y_t$ drawn from the denoising distribution $q(y_t \mid y_{t+k})$. Its purpose is to evaluate the authenticity of the generated samples and provide essential feedback to the generator, thereby improving its capability to produce realistic approximations of $y_t$.

The adversarial training process is formulated as a minimax optimization between the generator and the discriminator. The generator aims to produce samples that the discriminator cannot distinguish from real data, while the discriminator strives to accurately classify the generated samples as synthetic and the true samples as real. To stabilize the training and promote convergence, we incorporate a gradient penalty term into the discriminator's loss function. This penalty enforces smoothness in the discriminator's output with respect to its input. The gradient penalty is formulated following techniques proposed in recent studies~\cite{gungor2023adaptive,chan2024tutorial}, thereby bolstering learning efficiency and improving the model's robustness.

\begin{equation}\label{eq:discloss_01}
	\begin{aligned}
		L_D = & \sum_{t \ge 0} (\mathbb{E}_{q(y_0,y_t)} \mathbb{E}_{q(y_{t+k} \mid y_t)} \left[ -\log\left(\mathcal{D}_\theta(y_t, y_{t+k})\right) \right]  \\
		& + \mathbb{E}_{q(y_{t+k})} \mathbb{E}_{\mathcal{N}_\varphi(\mu_\varphi, \sigma_\varphi)} \left[ -\log(1-\mathcal{D}_\theta(\hat{y}_t, y_{t+k})) \right] \\
		& + \mathbb{E}_{q(y_0,y_t)} \mathbb{E}_{q(y_{t+k} \mid y_t)} \left[ \dfrac{1}{2} \parallel \nabla_{y_t}  \mathcal{D}_\theta (\hat{y}_t, y_{t+k}) \parallel_2^2 \right] )
		\end{aligned}
\end{equation}
where $ \mathcal{N}_\varphi(\mu_\varphi, \sigma_\varphi) $ is our generator parameterized by $ \varphi $ to reconstruct mean and variance. The generator loss becomes:
\begin{equation}\label{eq:loss_gen_01}
	\centering
	L_G = \sum_{t \ge 0} \mathbb{E}_{q(y_{t+k})} \mathbb{E}_{\mathcal{N}_\varphi(\mu_\varphi, \sigma_\varphi)} \left[ -\log(\mathcal{D}_\theta(y_t, y_{t+k})) \right]
\end{equation}

\subsection{Proposed self-supervised framework and training details}
Acquiring fully sampled data can be impractical due to constraints such as voluntary and involuntary motions, lengthy acquisition times, and signal decay. These constraints hinder the application of supervised DL-based CS-MRI approaches. Thus, we proposed a self-supervised approach that randomly divided the sampling pattern $ \Omega $ into two sets $ \aleph $ and $ \Upsilon $ as given in (\ref{eq:ss_patterns_01}). These two sets have no elements in common except the center of \textit{k}-space.

\begin{equation}\label{eq:ss_patterns_01}
	\centering
	\Omega = \aleph \vee \Upsilon
\end{equation}

We used an under-sampling pattern $ \aleph $ to train our proposed model and define our DC layer as follows.

\begin{equation}
	\centering
	\hat{Y}_t^{\text{SSAD-MRI}} (k) = \left\{ \begin{array}{c} \hat{Y}_t^{\text{SSAD-MRI}} (k), \,\,\,\, \text{if} \,\,\,\,k \in \aleph \\ X(k), \,\,\,\,\,\,\,\,\,\,\,\text{if}\,\,\,\, k \notin \aleph  \end{array} \right.
\end{equation}
where capital letters refer to the Fourier transform of the corresponding parameters. In other words, our method updates the \textit{k}-space lines that were under-sampled and keeps the original \textit{k}-space lines that were not sampled during image acquisition.

Finally, our proposed method defined the loss function given in (\ref{eq:kl_loss_02}) using a sampling pattern $ \Upsilon $ as follows:

\begin{equation}\label{eq:kl_loss_03}
	\centering
	\text{\L{}}^\Upsilon =  \underset{\varphi}{\arg\min} \dfrac{1}{2 \sigma_q^2(t)}\dfrac{(1 - \alpha_t)^2}{(1 - \bar{\alpha}_t)\alpha_t}\left[\parallel \epsilon - \hat{\epsilon}_\varphi^\Upsilon (y_t) \parallel_2^2\right]
\end{equation}

\begin{figure}[tbh!]
	\centering
	\includegraphics[width=\textwidth, draft=false]{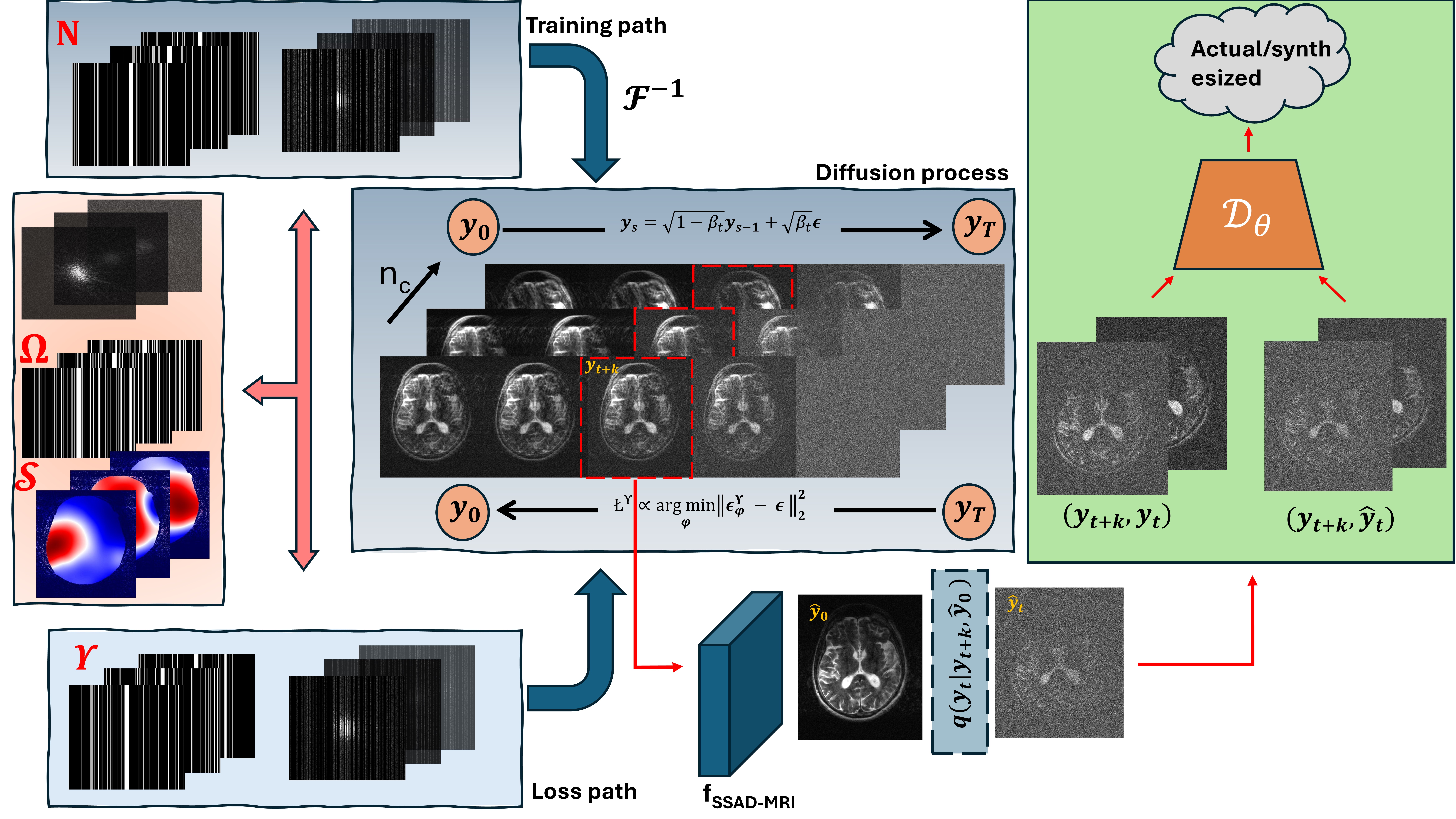}
	\caption{The flowchart of our proposed self-supervised adaptive diffusion model is illustrated. The coil sensitivity $ \mathcal{S} $, random sampling mask $ \Omega $, and \textit{k}-space were illustrated as inputs. Firstly, the random sampling pattern $ \Omega $ is divided into two non-overlapping sampling masks $ \aleph $ and $ \Upsilon $ that were used in the training path and loss path as given in (\ref{eq:kl_loss_03}), respectively. Then, the adversarial mapper was trained as given in (\ref{eq:discloss_01}) and (\ref{eq:loss_gen_01}) using the data sampled in step $ t+k $. The sample $ y_{t+k} $ used to recover $ \hat{y}_0 $ where was used to calculate $ \hat{y}_t $ in a given step $ t $ Equation (\ref{eq:sampling_qt_from_q0}).}
	\label{fig:flowchart}
\end{figure}

The final loss function is composed of discriminator loss, generator loss, and reconstruction loss as follows:

\begin{equation}\label{eq:all_loss_01}
	L_{\text{final}} = \text{\L{}}^\Upsilon + \lambda (L_D + L_G)
\end{equation}
where $ \lambda =0.1$ was used to scale and control the ratio between the losses. Our proposed method used the indices specified by $ \aleph $ to reconstruct \textit{k}-space at indices given by $ \Upsilon $. Figure~\ref{fig:flowchart} illustrates the flowchart of our proposed method. Then the loss function was calculated at the location indicated by the $ \Upsilon $ pattern. In other words, our self-supervised network was trained to decrease the discrepancy between the predicted images $ y^\aleph $ and the acquired measurement $ y^\Upsilon $ that was not seen in the training. At the inference step, the test unseen under -sampled data using a sampling pattern $ \Omega $ was used to reconstruct the fully sampled data.

This self-supervised scenario is similar to the cross-validation concept used to reduce bias and the likelihood of overfitting. Cross-validation partitions the dataset into at least two sets where one set is used to train a model and another set is used to evaluate the model. However, unlike cross-validation in machine learning, which performs partitioning once for all data, our method performs random partitioning per image slice.

In the inference step, we did not start from complete noise. Instead, we set the Gaussian noise covariance to 0.1 and then used them directly as initial inputs. This approach has been previously employed to reconstruct high-resolution MR images from under-sampled \textit{k}-space data and remove MRI motion artifacts~\cite{huang2023cdiffmr,safari2024mri}.

We used the original implementation of the recently proposed transformer, named ReconFormer\footnote{\url{https://github.com/guopengf/ReconFormer}}, as a generator. It incorporates the pyramid structures, enabling scale processing at each pyramid unit, while the globally columnar structure maintains high-resolution information~\cite{guo2023reconformer}. The discriminator consisted of four convolution layers with kernel size $ (3 \times 3) $ and padding one. Each convolution layer was followed by a ReLu activation layer and a batch normalization layer.

Our proposed method was trained using the Adam optimizer with a learning rate of $2\times 10^{-4}$ to minimize the loss function, employing a batch size of four over 25 epochs. Training was performed on our server using a single NVIDIA A100 GPU with CUDA Toolkit 12.2 and the PyTorch framework version 2.1.2~\cite{paszke2019pytorch}. The total training time was approximately 175 hours. The inference time for reconstructing a single image slice was about 14 seconds. While we used an A100 GPU for training, the model can also be trained on GPUs with less memory, such as an NVIDIA RTX 3090, by adjusting the batch size to accommodate the available GPU memory.

\subsection{Sampling masks}
We employed a 1d random sampling pattern $ \Omega $ where the \textit{k}-space center was excluded from sampling.The center fraction excluded from sampling was set to $ 4\% $ of the \textit{k}-space lines in the horizontal direction. Our proposed SSAD-MRI divides the acquired sampling mask $ \Omega $ into two disjoint sets $ \aleph $ and $ \Upsilon $ with except four \textit{k}-space lines at the center of dataset. Figure S1 in the supplementary document presents the $ \Omega $, $ \aleph $, and $ \Upsilon $ sampling masks across different values of $ \rho \in {0.3, 0.5, 0.7}$ with a fixed acceleration rate of $ R = 4\times $.

The sampling pattern $ \Upsilon $ was randomly sampled for each different slice to train the model. Therefore, our subsampled data using $ \Upsilon $ will be able to simulate the ghosting that is present in zero-filled datasets retrospectively subsampled using $ \Omega $ masks. We investigated a uniformly random selection among elements of $ \Omega $ to create subsampled patterns $ \aleph $ and $ \Upsilon $. The sampling ratio $ \rho = \frac{\mid \aleph \mid}{\mid \Upsilon \mid} \in \{0.3, 0.5, 0.7\}$ was used to train and test the proposed model, where $ \mid \bullet \mid $ is the number of elements. In other words, $ \rho = 0.3 $ means 30\% of $ \Omega $ was randomly assigned to $ \aleph $ to train the model and the rest were assigned to $ \Upsilon $ sampling pattern to define loss function as given in Equation (\ref{eq:kl_loss_03}).

\subsection{Dataset}\label{sec:data}
We utilized publicly available single-coil and multi-coil brain MRI datasets. Both datasets were retrospectively under-sampled using a random sample provided in the fastMRI database, with acceleration rates (R) equal to 2, 6, and 8~\cite{zbontar2018fastMRI,knoll2020fastmri}.

\subsubsection{Single-coil data:}

The brain MPI-Leipzig Mind-Brain-Body dataset~\cite{ds000221_00002, babayan2019mind} was used to train and test our proposed model. We utilized high-resolution MP2RAGE T1 maps of 318 patients that were split into two non-overlapping sets: training (250 patients) and testing (68 patients). The sagittal acquisition orientation of volumetric MP2RAGE \ce{T1} maps with 176 slices were acquired with the imaging parameters: \ce{T_R} = 5000 ms, \ce{T_R} = 2.92 ms, \ce{T_R} = 700 ms, \ce{TI2} = 2500 ms, \ce{FA1} = 4$^\circ$, \ce{FA2} = 5$^\circ$, pre-scan normalization, echo spacing = 6.9 ms, bandwidth = 240 Hz/pixel, FOV = 256 mm, voxel size = 1 mm isotropic, GRAPPA acceleration factor = 3, slice order = interleaved, duration = 8 min 22 s.

\subsubsection{Multi-coil data:}
The brain datasets used in the preparation of this article were obtained from the NYU fastMRI Initiative database\footnote{\url{fastmri.med.nyu.edu}} that was approved by the NYU School of Medicine Institutional Review Board~\cite{zbontar2018fastMRI,knoll2020fastmri}. We utilized \ce{T2}-weighted (\ce{T2}-w) images of 1051 patients' data for training and 325 patients' unseen data for testing. In addition, we evaluated the robustness of our model to domain shift using two out-of-distribution (OOD) unseen datasets, \ce{T1}-weighted (\ce{T1}-w)  and postcontrast \ce{T1}-w (T1c), using 50 patients' data for each MRI sequence. 

The sensitivity maps $ \mathcal{S} $ were generated from $ 24 \times 24 $ center of \textit{k}-space using ESPIRiT~\footnote{\url{https://mrirecon.github.io/bart/}}~\cite{uecker2014espirit} with a kernel size $ 6 \times 6 $ and the calibration matrix and eigenvalue decomposition threshold 0.02 and 0.95, respectively.

\subsection{Quantitative analysis}
To evaluate the performance of the proposed SSAD-MRI model, we compared it against two benchmark models: SS-MRI~\cite{hu2021self} and ReconFormer~\cite{guo2023reconformer}, which were trained under our proposed sampling approach. 

Three quantitative metrics were employed: normalized mean square error (NMSE), peak signal-to-noise ratio (PSNR), and structural similarity index (SSIM), utilizing the PyTorch Image Quality library~\footnote{\url{https://piq.readthedocs.io/en/latest/index.html}}~\cite{kastryulin2022piq}. The NMSE compared reconstructed $ \hat{y} $ with ground truth $ y $ as follows:

\begin{equation}\label{eq:nmse_eq01}
	\centering
	\text{NMSE}(y, \hat{y}) = \dfrac{\parallel y - \hat{y} \parallel_2^2}{\parallel y \parallel_2^2}
\end{equation}
where $ \parallel \bullet \parallel_2 $ is the squared Euclidean distance. Lower NMSE values indicate better image reconstruction. However, it might be in favor of blurry images~\cite{zbontar2018fastMRI}.

The PSNR defined below utilizes a logarithmic scaling that makes the quantification results more aligned with human perception~\cite{eidex2024high}. 

\begin{equation}\label{eq:psnr_eq01}
	\centering
	\text{PSNR}(y, \hat{y}) = \dfrac{y_{\max}^2}{\frac{1}{N}\sum_{i=1}^{N} (y_i - \hat{y}_i)^2}
\end{equation}
where $ y_{\max} $ is the maximum signal intensity of ground truth images. Higher PSNR indicates better image reconstruction.

The SSIM quantifies the structural similarity between the reconstructed $ \hat{y} $ and ground truth $ y $ is defined as follows:

\begin{equation}\label{eq:ssim_eq01}
	\centering
	\text{SSIM}(y, \hat{y}) = \dfrac{(2\mu_y \mu_{\hat{y}} + c_1) (2\sigma_{y\hat{y}} + c_2) }{(\mu_y^2 + \mu_{\hat{y}}^2 + c_1) (\sigma_{y}^2 + \sigma_{\hat{y}}^2 + c_2)}
\end{equation}
where $ \mu_y $ and $ \mu_{\hat{y}} $ are the average voxel values of $ y $ and $ \hat{y} $ images, $ \sigma_{y} $ $ \sigma_{\hat{y}} $ are the variance, and $ \sigma_{y\hat{y}} $ is the covariance between $ y $ and $ \hat{y} $ images. The constants $ c_1 $ and $ c_2 $ stabilize the division, we used $ c_1 = 0.01 y_{\max} $ and $ c_2 = 0.03 y_{\max} $. SSIM ranges between -1 and +1, with the best similarity achieved by an SSIM equal to +1.

\subsection{Statistical analysis}\label{sec:evaluation_methods}

The quantitative metrics were compared using one-way analysis of variance (ANOVA) to evaluate the null hypothesis that the mean values of each method were the same. The differences with \textit{p} $< 0.05$ was considered statistically significant. Additionally, a multi-comparison Tukey’s honestly test difference (HSD) was performed to evaluate pairwise differences between the methods, with \textit{p} $< 0.05$ indicating statistical significance.

We reported the average values of the quantitative metrics. In addition, we calculated the $ 95\% $ confidence intervals (CIs) on the average values using the percentile bootstrap method (with $ n=10000 $ iterations) with the bias-adjusted and accelerated bootstrap method~\cite{efron1993confidence}.

\section{Results}\label{sec:result}
In this section, we present the comprehensive evaluation of our proposed SSAD-MRI model for MRI reconstruction. The results are organized to highlight the model's performance across different datasets and under various conditions.

We assess the within-domain multi-coil \ce{T2}-w reconstruction and single-coil MP2RAGE \ce{T1} map reconstruction capabilities of SSAD-MRI using both qualitative and quantitative metrics. The model's ability to preserve fine structures and reduce artifacts is demonstrated through visual comparisons and statistical analysis. The impact of loss mask partitioning on the performance of the proposed method is demonstrated.

We also evaluate the robustness of SSAD-MRI against domain shifts using OOD datasets. This analysis illustrates the model's adaptability to new and unseen data.

\subsection{Within-domain reconstruction}\label{sub:within_domain_reconstruction}
Our proposed SSAD-MRI was evaluated for within-domain reconstruction at $ R=2\times $, $ 4\times $, and $ 8\times $ with a partitioning rate $ \rho=0.5 $. We conducted both qualitative and quantitative comparisons with the ReconFormer Transformer and SS-MRI models.

\subsubsection{Qualitative Results}
\begin{figure}[tbhp]
	\centering
	\includegraphics[width=\textwidth, draft=false]{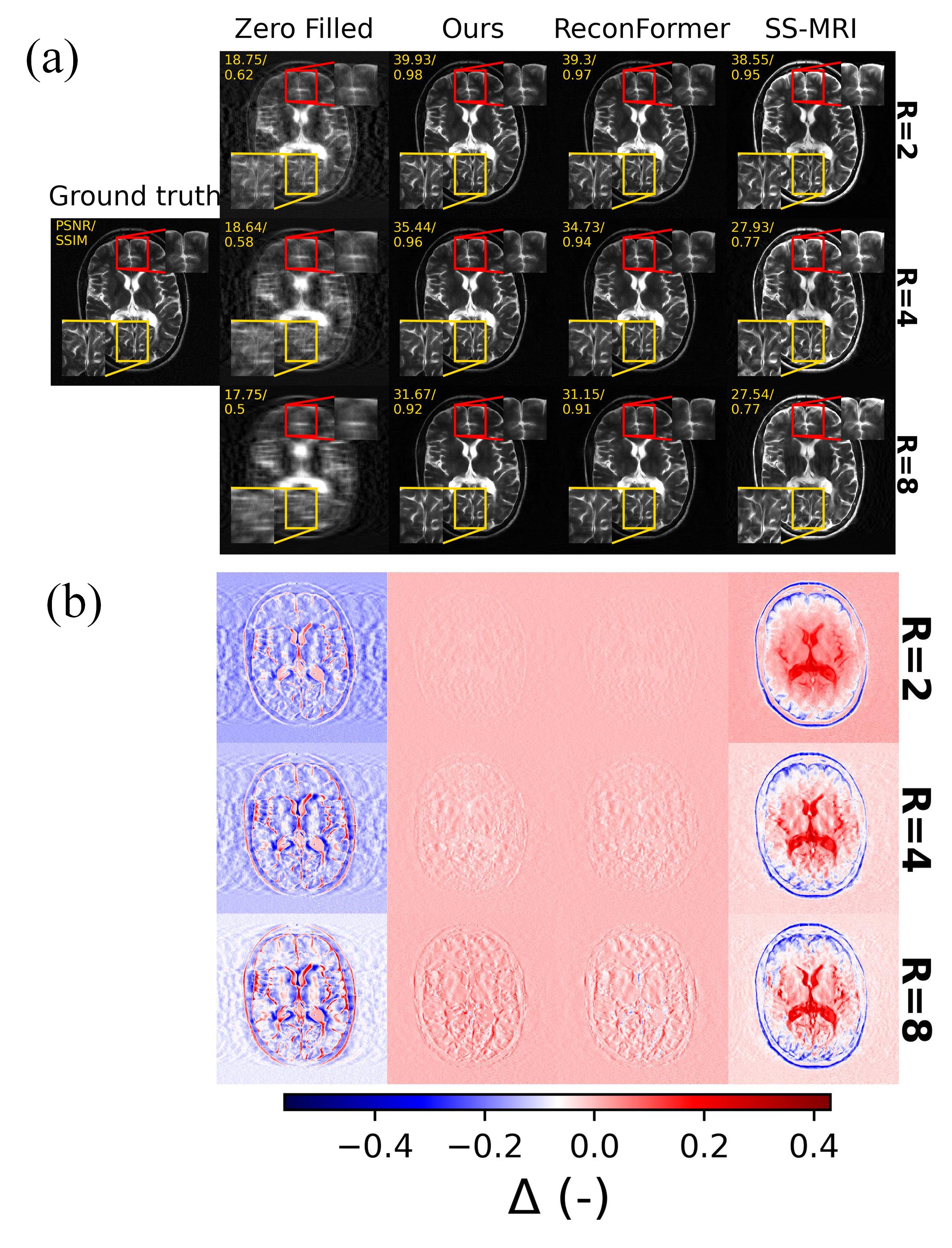}
	\caption{Within-domain axial \ce{T2}-w image reconstruction at $ R=2\times $, $ 4\times $, and $ 8\times $ and $ \rho=0.5 $ are illustrated. (a) illustrates the results for a healthy subject and (b) illustrates the difference map between the reconstructed and ground truth images.}
	\label{fig:mutli_coil_healthy_volunteer}
\end{figure}

\paragraph{Multi-coil dataset:}
Figures~\ref{fig:mutli_coil_healthy_volunteer} and S2 present the qualitative results of within-domain \ce{T2}-w multi-coil images for two subjects. Figure~\ref{fig:mutli_coil_healthy_volunteer}(a) shows results for a healthy volunteer, highlighting that our method recovers more details, particularly in the gold boxed regions, compared to the ReconFormer model trained with our framework. Difference maps in Figure~\ref{fig:mutli_coil_healthy_volunteer}(b) illustrate that our method better preserves gray matter and edges, indicated by the gold and red boxes.

Figure S2 displays results for a patient with brain abnormalities. Our proposed method could recover the abnormality boundary shown by a red box close to the ground truth. Furthermore, our method reconstructed details better (gold boxes) compared with the SS-MRI method. 

\begin{figure}[tbhp]
	\centering
	\includegraphics[width=.9\textwidth, draft=false]{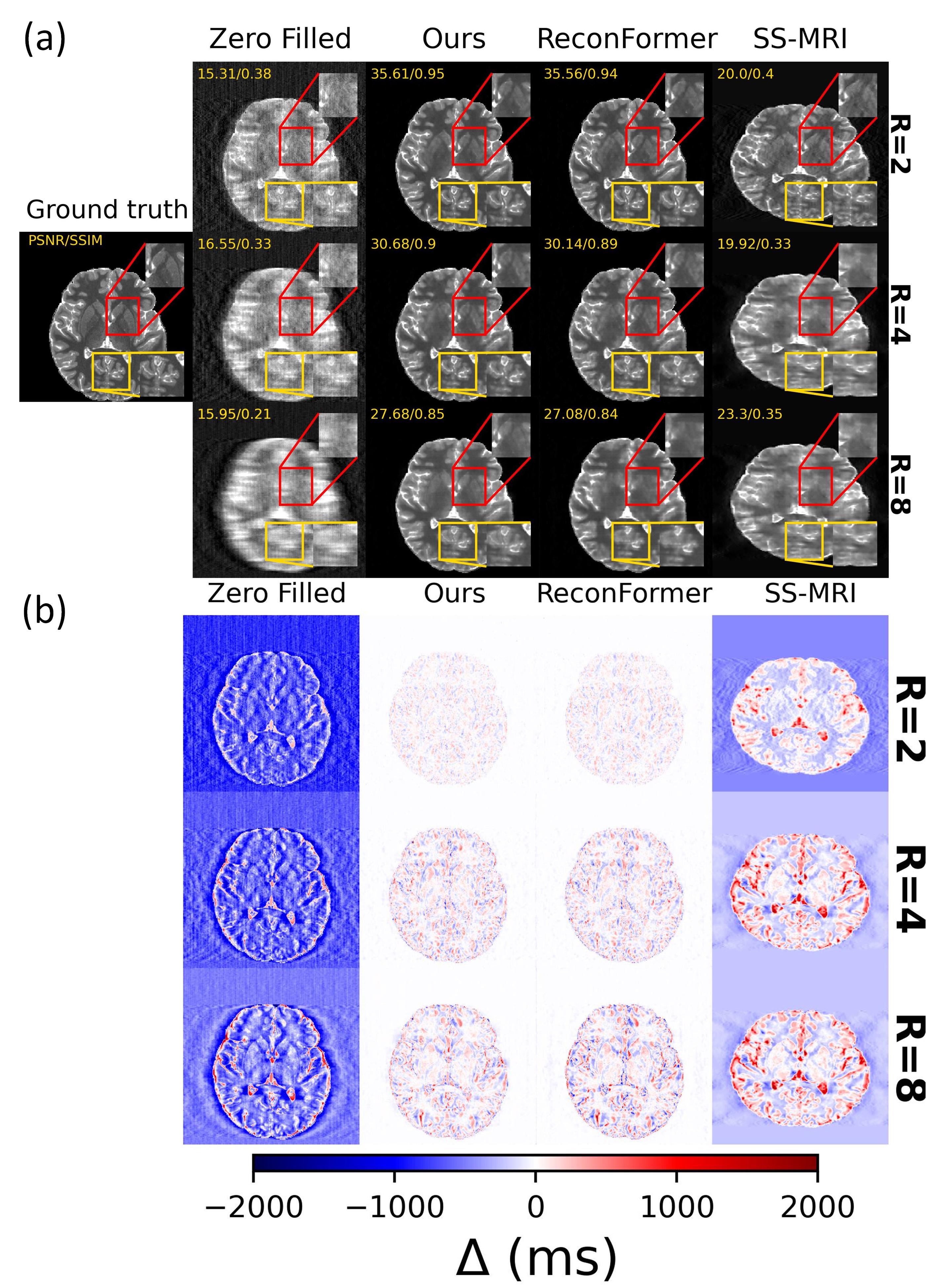}
	\caption{Reconstruction of within-domain MP2RAGE \ce{T1} quantitative maps at acceleration factors $ R \in \{ 2\times, 4\times, 8\times\} $ with $ \rho=0.5 $. (a) Reconstructed images for a representative subject. (b) Corresponding pixel-wise difference maps between the reconstructed images and the ground truth images.}
	\label{fig:single_coil_healthy_volunteer}
\end{figure}

\paragraph{Single-coil dataset:}
Figure~\ref{fig:single_coil_healthy_volunteer} presents the qualitative results for high-resolution single-coil MP2RAGE \ce{T1} quantitative map for a healthy volunteer at $ R=2\times $, $ 4\times $, and $ 8 \times $ with a partitioning rate $ \rho=0.5 $. Our method could preserve the small structure shown by gold boxes as well as the Putamen and Caudate nuclei shown by red boxes with better spatial contrast than the comparative models, as shown in Figure~\ref{fig:single_coil_healthy_volunteer}(a) and the difference map shown in Figure~\ref{fig:single_coil_healthy_volunteer}(b). 

\subsubsection{Quantitative Results}
Tables \ref{tab:multicoil_withinDoamin} and \ref{tab:singlecoil_withinDoamin} summarize the quantitative metrics for multi-coil and single-coil datasets, respectively, at different acceleration rates. The ANOVA test indicated statistically significant differences (\textit{p} $ \ll 10 ^{-5} $) between the average values of the methods. The results of the Tukey's HSD multi-comparison are presented in the text and tables.

\paragraph{Multi-coil dataset:}
The quantitative results are listed in Table~\ref{tab:multicoil_withinDoamin} at $ R=2\times $, $ 4\times $, and $ 8 \times $ and $ \rho=0.5 $. The ANOVA test indicated \textit{p} $ \ll 10^{-5} $ for all metrics indicating that there are statistically significant differences between average values. Our method got the lowest NMSE for all acceleration rates except for $ R=2\times $ and $ R=4\times $. The ReconFormer method achieved the lowest NMSE value at $ R=2\times $, nonetheless, it was not statistically significantly different from our method (\textit{p} $ =0.65 $). Although our method achieved the lowest NMSE value for $ R=8\times $, but it was not  statistically significantly different from ReconFormer (\textit{p} $ = 0.83$). Our proposed self-supervised method achieved the highest PSNR and SSIM values for all acceleration rates, demonstrating the lowest remaining spatial distortion, like ghosting, and the highest structural similarities between the ground truth and reconstructed images, respectively.

\begin{table}[b]
	\caption{Within-domain performance for multi-coil axial \ce{T2}-w fastMRI at $ R\in \{ 2\times, 4\times, 8\times\} $ and $ \rho=0.5 $ are provided. The arrows indicate directions of better performance.}
	\label{tab:multicoil_withinDoamin}
	\resizebox{\columnwidth}{!}{%
		\begin{tabular}{llllll}
			\hline
			\rowcolor[HTML]{FFFFFF} 
			&    & Zero filled           & Reconformer           & SS-MRI                & Ours                  \\ \hline 
			\rowcolor[HTML]{FFFFFF} 
			\cellcolor[HTML]{FFFFFF}                                          & 2$ \times $ & 14.99 (12.95 - 17.14) & {0.47 (0.45 - 0.50)}$ ^* $   & 0.52(0.50 - 0.56)$ ^* $   & 0.51 (0.48 - 0.54)   \\
			\rowcolor[HTML]{FFFFFF} 
			\cellcolor[HTML]{FFFFFF}                                          & 4$ \times $ & 26.38 (24.43 - 28.41) & 1.51 (1.42 - 1.64)    & 4.87 (4.76 - 4.98)    & {1.26 (1.18 - 1.35)}    \\
			\rowcolor[HTML]{FFFFFF} 
			\multirow{-3}{*}{\cellcolor[HTML]{FFFFFF}NMSE (95\% CI) [\%]  $ \downarrow $}          & 8$ \times $ & 30.78 (28.49 - 33.11) & 3.34 (3.14 - 3.63)$ ^* $   & 5.16 (5.06 - 5.26)    & 3.26 (3.08 - 3.51)    \\
			\rowcolor[HTML]{EFEFEF} 
			\cellcolor[HTML]{EFEFEF}                                          & 2$ \times $ & 17.75 (17.64 - 17.86) & 38.80 (38.65 - 38.96) & 39.05 (38.90 - 39.20) & 39.93 (39.78 - 40.09) \\
			\rowcolor[HTML]{EFEFEF} 
			\cellcolor[HTML]{EFEFEF}                                          & 4$ \times $ & 17.64 (17.53 - 17.77) & 34.23 (34.10 - 34.37) & 27.93 (27.79 - 28.06) & 35.44 (35.31 - 35.58) \\
			\rowcolor[HTML]{EFEFEF} 
			\multirow{-3}{*}{\cellcolor[HTML]{EFEFEF}PSNR (95\% CI) [dB] $ \uparrow $}  & 8$ \times $ & 18.75 (18.62 - 18.87) & 27.54 (27.43 - 27.65) & 30.65 (30.52 - 30.76) & 31.67 (31.55 - 31.79) \\
			\rowcolor[HTML]{FFFFFF} 
			\cellcolor[HTML]{FFFFFF}                                          & 2$ \times $ & 81.53(81.06 - 81.98)  & 96.10(96.02 - 96.17)  & 96.33 (96.15 - 96.49) & 98.14 (98.06 - 98.21) \\
			\rowcolor[HTML]{FFFFFF} 
			\cellcolor[HTML]{FFFFFF}                                          & 4$ \times $ & 70.89(70.44 - 71.36)  & 93.36(93.19 - 93.51)  & 76.72 (76.32 - 77.15) & 95.55 (95.38 - 95.69) \\
			\rowcolor[HTML]{FFFFFF} 
			\multirow{-3}{*}{\cellcolor[HTML]{FFFFFF}SSIM (95\% CI) [\%] $ \uparrow $}   & 8$ \times $ & 65.17(64.65 - 65.68)  & 89.72 (89.49 - 89.93) & 77.01 (76.63 - 77.39) & 91.67 (91.45 - 91.89) \\ \hline
			\rowcolor[HTML]{FFFFFF} 
			\multicolumn{6}{l}{\cellcolor[HTML]{FFFFFF}$ ^* $ indicates \textit{p}-value $ > 0.05 $ of Tukey's HSD. }                                                                                                                  
		\end{tabular}%
	}
\end{table}

\paragraph{Single-coil dataset:}

The quantitative metrics for the within-domain single-coil high-resolution MP2RAGE T1 map are summarized in Table~\ref{tab:singlecoil_withinDoamin}. One-way ANOVA tests indicated significant differences (\textit{p} $ \ll 10^{-5} $) between methods for all metrics. Tukey’s HSD test confirmed that SSAD-MRI performed significantly better than comparative methods for most metrics. Our method achieved the lowest NMSE values that were statistically different from the comparative methods for all acceleration rates except with $ R=2 $ where our method achieved a performance that did differ statistically significant (\textit{p}-value $ =0.97 $) from ReconFormer. Our method achieved the highest PSNR values for all acceleration rates that were statistically significantly (\textit{p} $ <0.05 $) from comparative methods. Furthermore, our proposed method achieved the highest SSIM at $ R=4\times $ that differed statistically significantly higher (\textit{p} $ <0.05 $) than the other methods. Nonetheless, our method and ReconFormer performed similarly in terms of the SSIM index at $ R = 2\times $ and $ 8\times $ with \textit{p} of $ 0.15 $ and $ 0.61 $, respectively.

\begin{table}[tbhp!]
	\caption{Within-domain performance for single-coil high resolution MP2RAGE T1 map at $ R\in \{ 2\times, 4\times, 8\times\} $ and $ \rho=0.5 $ are provided. The arrows indicate directions of better performance.}
	\label{tab:singlecoil_withinDoamin}
	\resizebox{\columnwidth}{!}{%
		\begin{tabular}{llllll}
			\hline
			\rowcolor[HTML]{FFFFFF} 
			&    & Zero filled           & Reconformer            & SS-MRI                & Ours                                         \\ \hline
			\rowcolor[HTML]{FFFFFF} 
			\cellcolor[HTML]{FFFFFF}                                          & 2$ \times $ & 15.37 (13.46 - 17.49) & 0.53 (0.52 - 0.55)$ ^* $  & 5.18 (4.97 - 5.40)  & 0.56 (0.54 - 0.57)                           \\
			\rowcolor[HTML]{FFFFFF} 
			\cellcolor[HTML]{FFFFFF}                                          & 4$ \times $ & 22.43 (20.65 - 24.33) & 2.04 (1.99 - 2.09)     & 8.75 (8.36 - 9.18) & 1.87 (1.82 - 1.93)                           \\
			\rowcolor[HTML]{FFFFFF} 
			\multirow{-3}{*}{\cellcolor[HTML]{FFFFFF}NMSE (95\% CI) [\%] $ \downarrow $}          & 8$ \times $ & 33.17 (31.79 - 34.72) & 3.99 (3.90 - 4.10)     & 10.76 (10.42 - 11.13) & 3.65 (3.56 - 3.75)                           \\
			\rowcolor[HTML]{EFEFEF} 
			\cellcolor[HTML]{EFEFEF}                                          & 2$ \times $ & 15.46 (15.39 - 15.53) & 35.42(35.31 - 35.53)   & 23.07(22.98 - 23.16)  & 36.15(36.03 - 36.26)                         \\
			\rowcolor[HTML]{EFEFEF} 
			\cellcolor[HTML]{EFEFEF}                                          & 4$ \times $ & 15.35 (15.28 - 15.43) & 29.46(29.37 - 29.56)   & 21.31(21.21 - 21.40)  & 30.51(30.42 - 30.61)                         \\
			\rowcolor[HTML]{EFEFEF} 
			\multirow{-3}{*}{\cellcolor[HTML]{EFEFEF}PSNR (95\% CI) {[}dB{]} $ \uparrow $} & 8$ \times $ & 16.31 (16.23 - 16.40) & 26.54(26.45 - 26.65)   & 20.75(20.66 - 20.85)  & 27.94(27.84 - 28.04)                         \\
			\rowcolor[HTML]{FFFFFF} 
			\cellcolor[HTML]{FFFFFF}                                          & 2$ \times $ & 73.83(73.43 - 74.21)  & 95.75(95.68 - 95.81)$ ^* $ & 85.12(84.57 - 85.65)  & 95.31(95.25 - 95.38)                         \\
			\rowcolor[HTML]{FFFFFF} 
			\cellcolor[HTML]{FFFFFF}                                          & 4$ \times $ & 68.62(68.29 - 68.94)  & 89.11(88.96 - 89.26)   & 80.47(80.06 - 80.89)  & 89.80(89.63 - 89.95)                         \\
			\rowcolor[HTML]{FFFFFF} 
			\multirow{-3}{*}{\cellcolor[HTML]{FFFFFF}SSIM (95\% CI) {[}\%{]} $ \uparrow $} & 8$ \times $ & 66.13(65.83 - 66.42)  & 84.55(84.32 - 84.78)$ ^* $ & 77.17(76.75 - 77.62)  & \cellcolor[HTML]{FFFFFF}84.77(84.53 - 84.99) \\ \hline
			\rowcolor[HTML]{FFFFFF} 
			\multicolumn{6}{l}{\cellcolor[HTML]{FFFFFF}$ ^* $ indicates \textit{p}-value $ > 0.05 $ of Tukey's HSD.}                                                                                                                                          
		\end{tabular}%
	}
\end{table}

\subsubsection{Voxel-wise correlation}

We visualized the scatter plot for 80 randomly selected within-domain multi-coil and single-coil data to visualize and quantify the agreement between reconstructed images at $ R=8\times $ and $ \rho=0.5 $ with the ground truth reference images (see Figure S3). For the single-coil dataset, our method reconstructed images shown in Figure~S3(c) with better conformity with the references markedly better than SS-MRI shown in Figure S3(a) and comparable to ReconFormer shown in Figure~S3(b). The same results were achieved for multi-coil data where the SS-MRI performed better than itself trained using the single-coil dataset. Still there is a noticeable gap between our proposed (see Figure~S3(f)) method and SS-MRI (see Figure~S3(d)). Furthermore, ground truth voxel-wise shown in Figure~S3(e) and (f) confirms that our method might be able to generate images that are more similar to the ground truth.

\subsubsection{Effect of Sampling Mask Ratio}
Our proposed method split the sampling mask $ \Omega $ randomly into two non-overlapping masks $ \aleph $ and $ \Upsilon $, which were used in train and loss paths, respectively (see Figure~\ref{fig:flowchart}). The ratio $ \rho = \frac{\mid \aleph \mid}{\mid \Upsilon \mid} $ plays a role in the network's performance. We trained and tested the network for varying $ \rho \in \{0.3, 0.5, 0.7\} $ (see Figure S1 in the supplementary for the sampling masks). The effect of network training with different $ \rho $s on the quantitative metrics NMSE, PSNR, and SSIM are shown in Figure~S4 at three acceleration rates $ R \in \{2, 4, 8\} $ that were acquired using multi-coil axial T2-w images. Our method achieved better performance at $ \rho = 0.5 $ in terms of PSNR and SSIM at $ R=2\times  $ and $ R=4 \times $. However, it performed similarly to $ \rho = 0.7 $ at $ R = 8\times $. Furthermore, NMSE metrics indicate that our method achieved the best performance when trained using $ \rho = 0.5 $ than other splitting ratios $ \rho \in \{0.3, 0.7\} $.

\subsection{Out-of-domain reconstruction}\label{sub:ood_reconstruction}

\begin{figure}[tbhp!]
	\centering
	\includegraphics[width=\textwidth]{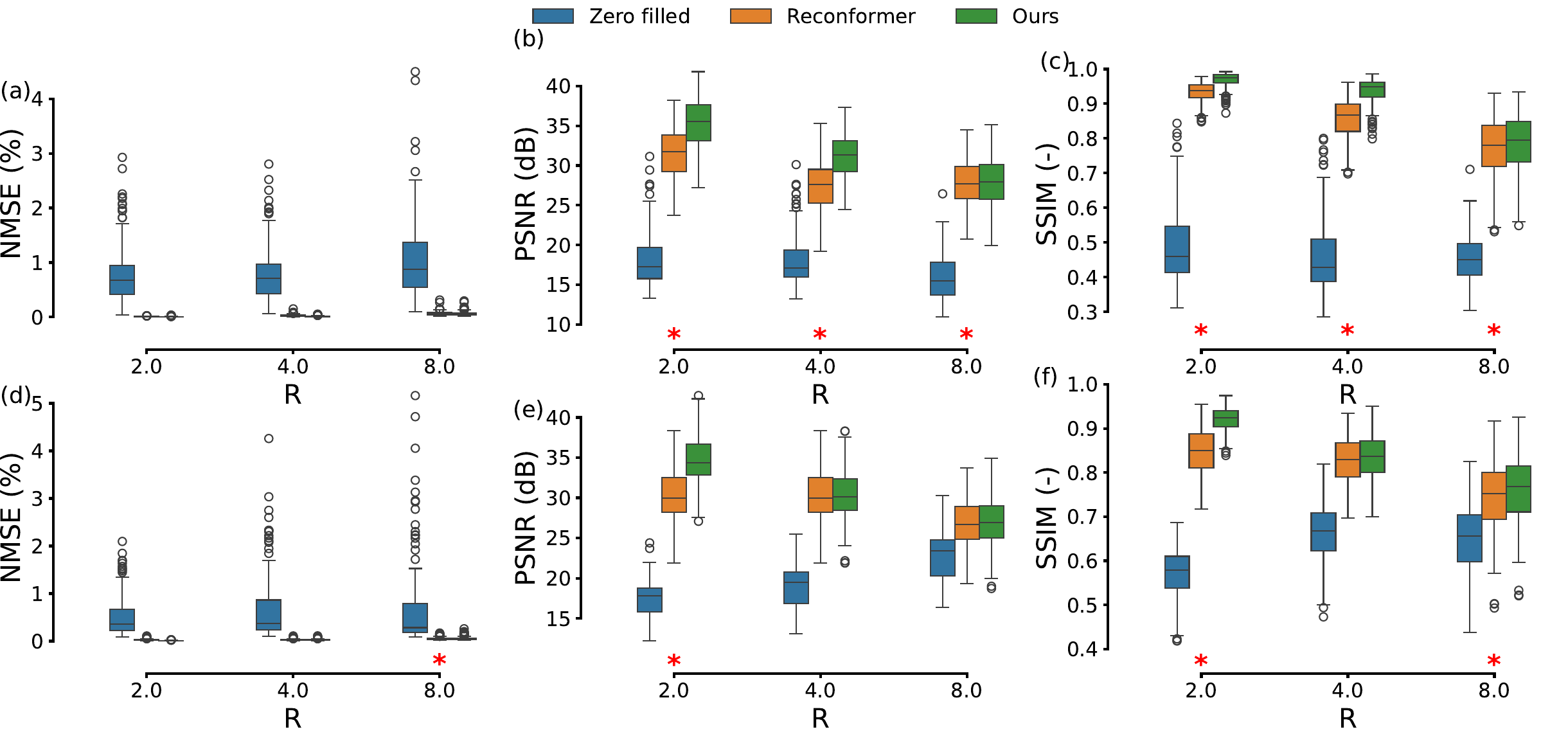}
	\caption{Out-of-distribution reconstruction quantitative results are illustrated for multi-coil axial T1-w and T1c in the first and second rows, respectively, for three difference acceleration rates. The red stars indicate statistically significant differences (\textit{p} $< 0.05$).}
	\label{fig:ood_qmetrics}
\end{figure}

We evaluated our model performance in OOD reconstructions where our proposed method was trained using multi-coil axial \ce{T2}-w fastMRI images and then tested using axial \ce{T1}c and \ce{T1}-w shown, respectively, in first and second rows of Figure~\ref{fig:ood_qmetrics} at $ R\in \{2\times, 4\times, 8\times\} $. Our method statistically significant (\textit{p} $ \ll 10^{-5} $) improved quantitative metrics. 

Our method achieved significantly lower NMSE values (\textit{p} $ \ll 10^{-5} $), indicating high voxel-wise similarity between the reconstructed and ground truth images for \ce{T1}-w and \ce{T1}c illustrated in Figures~\ref{fig:ood_qmetrics}(a) and (d), respectively. In addition, high PSNR values confirm minimal remaining spatial distortion in the reconstructed images for \ce{T1}-w and \ce{T1}c illustrated in Figures~\ref{fig:ood_qmetrics}(b) and (e), respectively. Finally, The high SSIM values demonstrate a remarkable resemblance between the reconstructed and ground truth images (see Figures~\ref{fig:ood_qmetrics}(c) and (f)).

\section{Discussion}\label{sec:discussion}

This study introduces the Self-supervised Adversarial Diffusion (SSAD-MRI) model, a novel self-supervised deep learning-based compressed sensing MRI (DL-based CS-MRI) method designed to mitigate the challenges associated with prolonged MRI acquisition times and the necessity for fully sampled datasets. Our approach leverages the synergy between CS and DL to accelerate MRI acquisition without compromising image quality.

The SSAD-MRI model's primary innovation lies in its use of an adversarial mapper within a self-supervised framework, eliminating the dependence on fully sampled training datasets. This advancement is particularly significant for clinical scenarios where acquiring fully sampled data is impractical. By integrating a diffusion model, our method enhances sampling efficiency and reconstruction quality. The backward diffusion process, executed in smaller steps, contributes to robust and efficient image sampling. Extensive testing on OOD datasets demonstrated significant improvements in NMSE, PSNR, and SSIM metrics. These results highlight the model's ability to generalize across various MRI sequences and patient-specific conditions, underscoring its versatility for clinical use.

The majority of DL-based CS-MRI methods use supervised learning to train networks~\cite{safari2024fastmri, terpstra2023accelerated, schlemper2017deep, guo2023reconformer, huang2023cdiffmr}. However, acquiring fully sampled can be challenging in some practical applications due to the long acquisition time, physiological constraints, and signal decay. For instance, fully sampled high-resolution brain MP2RAGE \ce{T1} maps can take around 24 minutes to acquire, which is impractical for large-scale studies and may lead to patient discomfort unless using acceleration approaches. Such a long acquisition time increases the likelihood of patient movements, which could substantially reduce the image quality. Thus, being able to use self-supervised DL-based CS MRI approaches is imperative to broaden their applications where acquiring such data are challenging. By integrating a diffusion model, our method enhances reconstruction quality. The backward diffusion process, executed in smaller steps, contributes to robust and efficient image sampling. Extensive testing on within domain and OOD datasets demonstrated significant improvements in NMSE, PSNR, and SSIM metrics. These results highlight the model's ability to generalize across various MRI sequences and patient-specific conditions, underscoring its versatility for clinical use.

Several self-supervised studies have been proposed to train models without using fully sampled data. For instance, a data-driven method of de-aliasing was proposed for single-coil data that performed an image-to-image translation~\cite{senouf2019self}. However, it did not encode the operator and used similar sampling patterns for training and loss that increased noise during test time. Alternatively, a self-supervised study was proposed, assuming data acquired using two different sampling patterns~\cite{liu2020rare}. Furthermore, physics-driven methods that unroll the training process were also proposed~\cite{yaman2020self}. However, the unrolling nature of the method might increase the test burden and time.

Our self-supervised method uses a DC layer that was trained end-to-end and utilizes the sampling pattern that is feasible to implement clinically. Our experiment using the multi-coil and single-coil datasets enabled us to recover target ground truth images at markedly lower acquisition times (see Figures~\ref{fig:mutli_coil_healthy_volunteer} - \ref{fig:single_coil_healthy_volunteer}). Specifically, our method could recover fine details highlighted by gold and red boxes better than the comparative models. The negligible amount of remaining aliasing in the reconstructed images was confirmed by Figures~\ref{fig:mutli_coil_healthy_volunteer} - \ref{fig:single_coil_healthy_volunteer}. Furthermore, this observation was confirmed by the low NMSE and high PSNR achieved in Tables~\ref{tab:multicoil_withinDoamin} and \ref{tab:singlecoil_withinDoamin}.

Although our method could achieve the best performance for $ \rho=0.5 $, Figure~S4 suggests that our method might be resilient to variations in the hyper-parameter $ \rho $ around $ \rho = 0.5 $. In addition, out-of-domain reconstruction indicated in Figure~\ref{fig:ood_qmetrics} with significantly better performance (p $ \ll 10^{-5} $) indicates reasonable robustness against domain shifts. However, the results indicate that our method performed better on T1-w (first row in Figure~\ref{fig:ood_qmetrics}) images than when tested on T1c (second row in Figure~\ref{fig:ood_qmetrics}). That might be due to contrast agent enhancement in T1c in the image regions that are not evident in T2-w images.

Despite the promising outcomes, several limitations need to be addressed. Our model was not tested on prospectively undersampled raw \textit{k}-space datasets acquired under parallel imaging frameworks. Future work should explore the application of SSAD-MRI to these datasets to validate its clinical utility further. In addition, we did not train our method using raw multi-coil high-resolution 3D MRI, such as \ce{T1} magnetization-prepared rapid acquisition gradient echoes MRI images because they are not readily available to the end-user. Expanding the training dataset to include such data could enhance the model's performance.

Addressing these limitations will enhance not only the model's performance but also its clinical applicability. By reducing acquisition times through high acceleration factors—potentially up to 8-fold. Integrating our method into existing clinical workflows is feasible because it does not require fully sampled training datasets; it can be incorporated into the reconstruction pipelines of current MRI systems without major alterations to imaging protocols. Focusing on model generalizability, computational efficiency, and adherence to regulatory standards will be crucial for clinical deployment. Collaboration between researchers, clinicians, and industry partners will facilitate the optimization of our method for image reconstruction and ensure its effectiveness across diverse clinical scenarios. Ultimately, this approach has the potential to lead to improved patient outcomes and enhanced diagnostic capabilities by providing high-quality images more efficiently.

\section{Conclusions}\label{sec:conclusion}

Our self-supervised adversarial diffusion model significantly improves the quality of reconstructed MRI images from undersampled data without requiring fully sampled training datasets, offering a promising solution for accelerating MRI acquisition. The proposed method has the potential to enhance clinical MRI practices by reducing scan times and improving image quality, which is crucial for accurate diagnosis and treatment planning. By decreasing imaging time and the likelihood of motion artifacts, our approach may benefit a wide range of applications where rapid and reliable MRI is essential.

\section*{Conflicts of interest}
There are no conflicts of interest declared by the authors.

\section*{Acknowledgement}
This research is supported in part by the National Institutes of Health under Award Number R56EB033332, R01EB032680, and R01CA272991.

\section*{Data availability}

Brain MRI data were obtained from the New York University fastMRI initiative database (\url{https://fastmri.med.nyu.edu/}) and the OpenNeuro MPI-Leipzig Mind-Brain-Body dataset (\url{https://openneuro.org/datasets/ds000221/versions/00002}). The datasets were acquired with the relevant institutional review board approvals.

%
%


\end{document}